\documentclass[
    ,final            
  ]
  {aipproc}

\layoutstyle{8x11single}

\begin{document}

\hfill{UCHEP-11-07
}

\title{$\Upsilon$(5S) Results from Belle: Strange Beauty and Other Beasts
\footnote{talk presented at the {\it  19$^{th}$ Particles \& Nuclei International Conference (PANIC)},  July 24-9, 2011}
}


\author{K. Kinoshita\\ (Belle Collaboration)}{
  address={University of Cincinnati,
PO Box 210011,
Cincinnati, OH 45221, USA}
}

\begin{abstract}
The B-factories have successfully exploited the unique advantages of the $\Upsilon$(4S) resonance to study many aspects of the $B_d$ and $B_u$ mesons.
The $\Upsilon$(10860) resonance, also known as $\Upsilon$(5S), is above mass threshold for the $B_s$ and shares many of the same advantages.  The Belle experiment has collected more than 120 fb$^{-1}$ at the $\Upsilon$(5S), corresponding to 7.9 million $B_s$ events.  First results from these data are presented, including the first observation of a baryonic decay of $B_s$, a novel measurement of $\sin 2\phi_1$, and observation of $h_b$(1P), $h_b$(2P), and two charged bottomonium-like states.
\end{abstract}

\maketitle


The $\Upsilon$(10860), ($Mc^2=10876\pm 11$~MeV/$c^2$,  $\Gamma=55\pm 28$~MeV)\cite{pdg}, is interpreted as $\Upsilon$(5S), the fourth excitation of the vector bound state of $b\bar b$, and is just above $B_s^*\bar B_s^*$ threshold.
The Belle experiment\cite{belle}, located at KEKB\cite{kekb}, although built primarily to measure  $CP$ asymmetries of $B$ meson decay in $e^+e^-$ annihilations at the $\Upsilon$(4S) resonance, collected in 2005-9 a total of 121.4~fb$^{-1}$ at the $\Upsilon$(5S) resonance, corresponding to 37 million resonance events and including 7.9 million $B_s$ events.
Part of this sample, 23.6~fb$^{-1}$, as well as an 8~fb$^{-1}$ nearby energy scan produced 11 publications covering several $B_s$ modes, $B$ events at the $\Upsilon$(5S), and bottomonium transitions and spectroscopy.
While rates of $B_s$ in $e^+e^-$ collisions are low compared to those for hadronic collisions,
$\sigma(e^+e^-\to \Upsilon$(5S))$\approx 0.3$~nb, the $\Upsilon$(5S) is nonetheless competitive for several aspects of $B_s$ decay; the $e^+e^-$ environment produces clean events, efficiently triggered,  with precisely known center-of-mass energy.
Furthermore, the $B$-factory offers an existing facility with high luminosity, a well-studied detector with precise and sensitive photon detection, and an abundance of $\Upsilon$(4S) data for comparisons.
The new preliminary results presented here, based on the full on-resonance set,  include $\bar{B}_s^0 \to \Lambda_c^+\pi^-\bar\Lambda$, a novel measurement of $\sin 2\phi_1\ (aka \sin 2\beta)$, and several new bottomonia.

The decay $\bar{B}_s^0 \to \Lambda_c^+\pi^-\bar\Lambda$ is reconstructed in the following sub-modes: $\bar\Lambda \to\bar p\pi^+$, $\Lambda_c^+\to p K^-\pi^+$ (inclusion of complex conjugate modes is implied).
The three types of $B_s$ events, $B_s \bar B_s$, $B_s \bar B_s^{*}/\bar B_s B_s^{*}$, and $B_s^{*} \bar B_s^{*}$ ,  are well identified and separated through ``full reconstruction'' of $B_s$ decays, where all decay products are measured, a method used with great success for non-strange $B$'s at the $\Upsilon$(4S).
Each candidate's energy and momentum in the $e^+e^-$ center-of-mass are evaluated as $\Delta E \equiv E_{cand}-E_{beam}$ and $M_{\rm bc}\equiv \sqrt{E_{beam}^2-p_{cand}^2}$.
The distribution of candidates in these two variables is fitted to extract the number of signal events, which is found to be $24\pm 7$ (5$\sigma$ significance).  
From this result we extract a  branching fraction
${\cal B}(\bar{B}_s^0 \to \Lambda_c^+\pi^-\bar\Lambda) = (4.8\pm 1.4(stat.)\pm 0.9 (sys)\pm 1.3 (\Lambda_c^+))\times 10^{-4}$
where the errors are statistical, systematic, and due to the uncertainty on the branching fraction of $\Lambda_c$ mode, respectively.
This value is consistent with that for the correponding channel for $B^-$, 
${\cal B}({B}^- \to \Lambda_c^+\pi^-\bar{p}) = (2.8\pm 0.8)\times 10^{-4}$\cite{pdg}.

\begin{figure}[h]
\includegraphics[width=5.2cm]{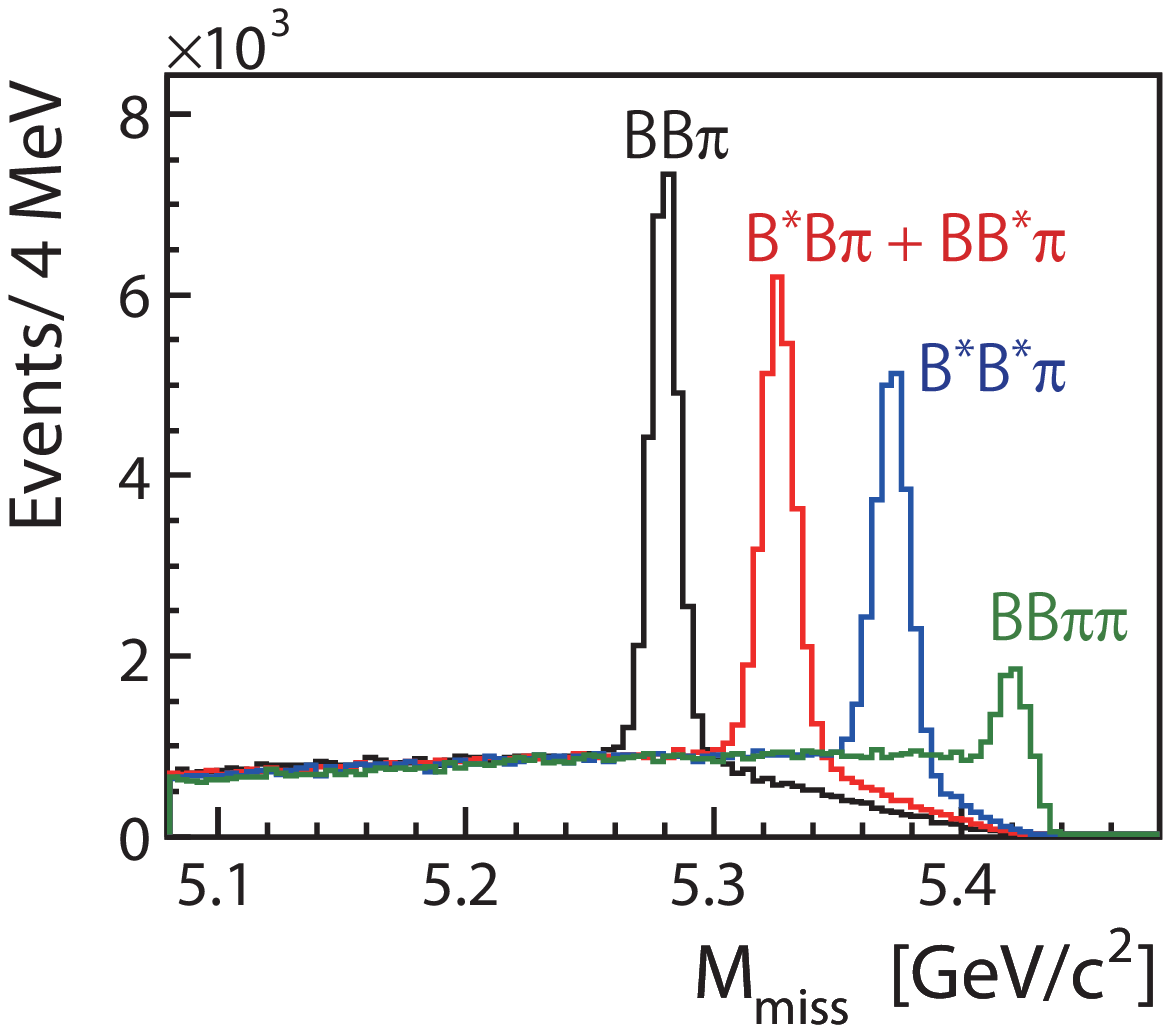}
\includegraphics[width=5.2cm]{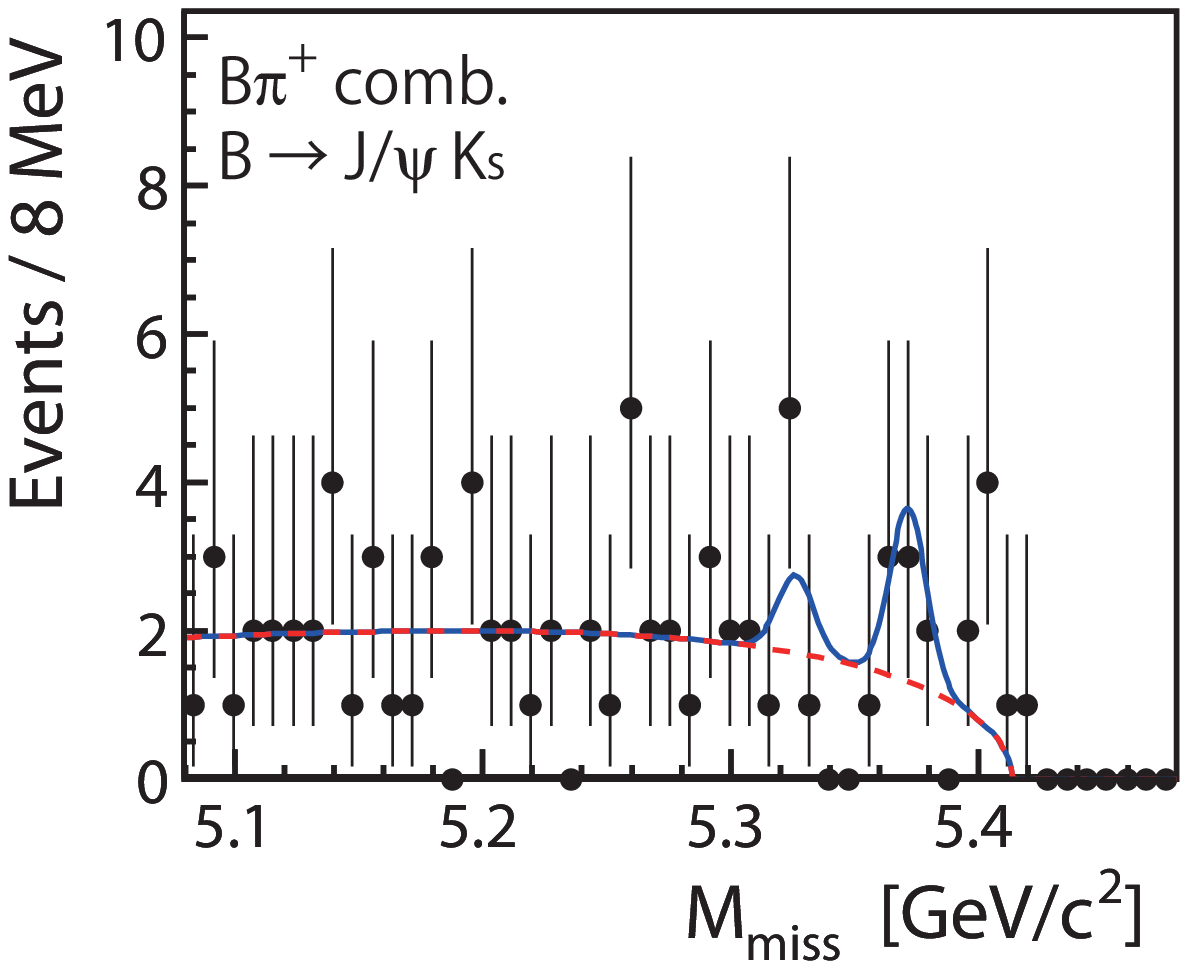}
\includegraphics[width=5.2cm]{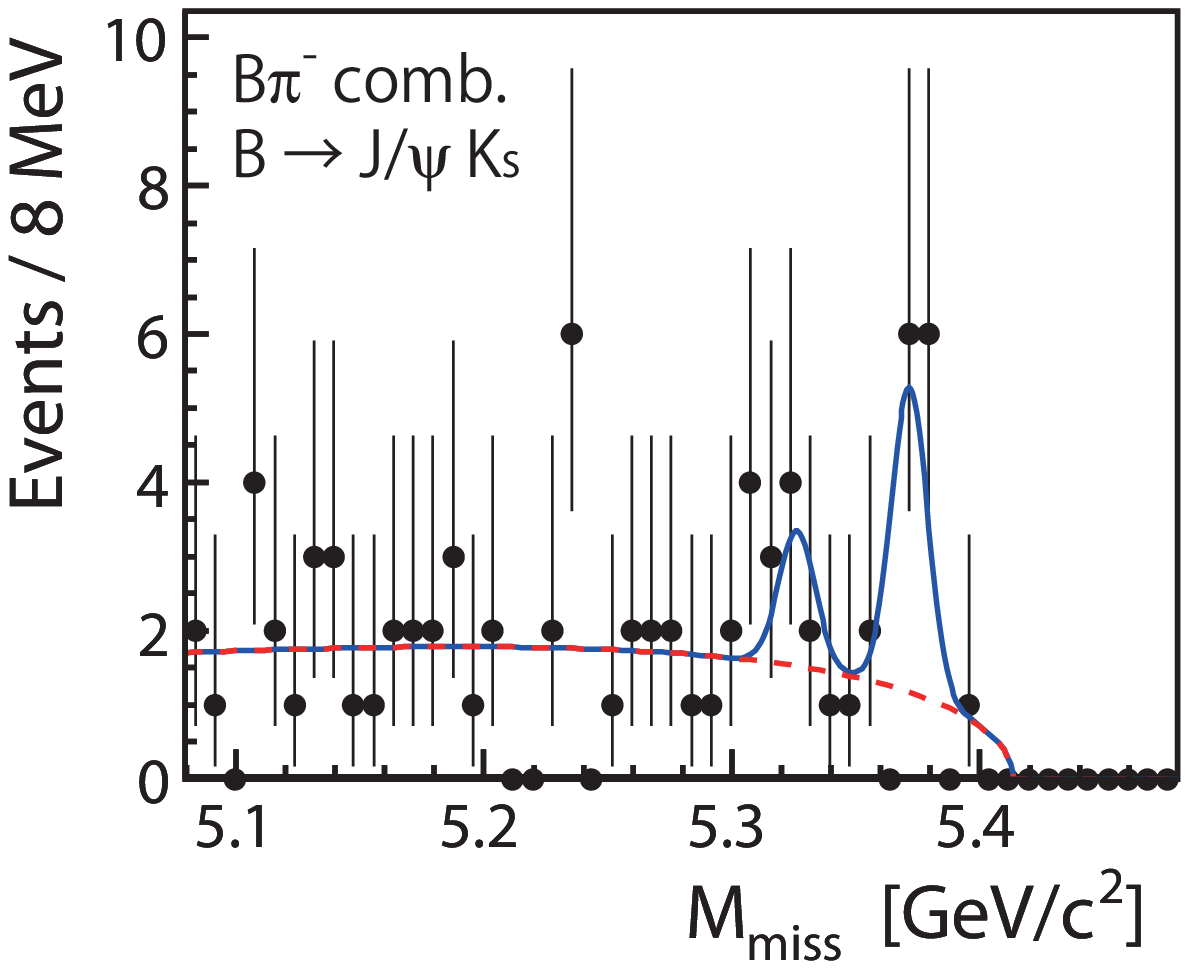}
\caption{Distributions in  $M_{miss}$ of tagged $B^0\pi$ candidates for (left) simulated $B\bar B\pi$, $B\bar B^*\pi$, and $B^*\bar B^*\pi$ events and 102.4~fb$^{-1}$ of data, (center) $B^0\pi^+$  and (right) $B^0\pi^-$.
\label{fig:BBpi}}
\end{figure}

While $B_s$ has been a primary focus of the $\Upsilon$(5S) program, the well-tuned methods of $B$ reconstruction from the $\Upsilon$(4S) have been applied to study the more complicated assortment of $B$ events at the $\Upsilon$(5S)\cite{BBpi}.
These methods are now applied to a novel tag that can be used to extract $\sin 2\phi_1$.
Three-body final states $B^{(*)0}\{\to B^0(\gamma)\} B^{(*)-}\pi^+$ (and charge conjugates) are identified through full reconstruction of a neutral $B$ in a $CP$-eigenstate and a charged pion.
The event residue, consisting of a charged $B$ and up to two photons, is characterized through ``missing mass,'' calculated through energy and momentum conservation:
\begin{eqnarray*}
E_{miss}=E_{beam}-E_{B^0\pi};\ \  \vec{p}_{miss}=-\vec{p}_{B^0\pi} ;\ \  MM(B^0\pi)=M_{miss}=\sqrt{E_{miss}^2-\vec{p}_{miss}^2}.
\end{eqnarray*}
The missing mass distributions are well separated for  $B\bar B\pi\pi$, $B\bar B\pi$, $B\bar B^*\pi$, and $B^*\bar B^*\pi$ events, as can be seen in Figure~\ref{fig:BBpi}(left).
The sign of the charged pion tags the initial flavor of the neutral $B$ and enables a {\it time-independent} measurement of $CP$-asymmetry, which is related to $\sin 2\phi_1$ as:
\begin{eqnarray*}
A_{BB\pi}\equiv \frac{N_{BB\pi^-}-N_{BB\pi^+}}{N_{BB\pi^-}+N_{BB\pi^+}}
=\frac{{\cal S}x+{\cal A}}{1+x^2}
\end{eqnarray*}
where ${\cal S}= -\eta_{CP}{\rm sin} 2\phi_1$ ($\eta_{CP}$ is the $CP$-eigenvalue of the $B^0$ mode), $x=\Delta m/\Gamma$, and ${\cal A}=0$ in the Standard Model.

Neutral $B$'s are reconstructed in the following modes and submodes:
$B^0\to J/\psi K_S;
\ J/\psi \to e^+e^-,\ \mu^+\mu^-$.
Figure~\ref{fig:BBpi} shows the distributions in $M_{miss}$ for (center) $B^0\pi^+$  and (right) $B^0\pi^-$  combinations, respectively, where the fits yield a total of $21.5\pm 6.8$ events.
The asymmetry is found to be $A_{BB\pi}=0.28\pm 0.28$, or $\sin 2\phi_1 = 0.57\pm 0.58\pm 0.06$.
This result establishes a new time-independent method of measuring $\sin 2\phi_1$, and the value is consistent with measurements in $\Upsilon$(4S) data.

The spectroscopy of bottomonia above $B\bar B$ threshold is not well known, and the data at and near the $\Upsilon$(5S) collected by Belle provide an excellent opportunity for exploration.
In addition to the fact that the predicted states $h_b$(1P,2P)($J^{PC}=1^{+-}$) and $\eta_b$(1S,2S,3S)($J^{PC}=0^{-+}$) are yet to be established, this activity is motivated by several recent findings:
the observation in the charm sector of many previously unpredicted charmonium-like states ($X,Y,Z$) by Babar, Belle, and CLEO; enhanced $h_c$ production in events $e^+e^-\to h_c\pi^+\pi^-$ in CLEO data at the mass of the charmonium-like state $Y(4260)$\cite{Yc_hc}; 
searches for $h_b$ at Babar (3.0$\sigma$ evidence \cite{babar_hb}) and CLEO (upper limit\cite{cleo_hb}) in $\Upsilon(3S)\to h_b\pi^0$; 
measurement at Belle of an anomalously high rate of $e^+e^-\to \Upsilon$(nS)$\pi^+\pi^-$ at $\Upsilon(5S)$ energy, roughly two orders of magnitude greater than at other $\Upsilon$'s\cite{Ypipi};
cross section for  $e^+e^-\to \Upsilon$(nS)$\pi^+\pi^-$ at CMS energies near the $\Upsilon$(5S) found to peak $\sim$20~MeV higher than the peak of the hadronic cross section\cite{5S_scan}.
In this talk I discuss the first observation of $h_b$(1S) and $h_b$(2S)\cite{hb_obs} as well as the first observation of a charged bottomonium-like state\cite{Zb_obs} that may be analogous to the $Z$(4430), a charmonium-like state seen at Belle\cite{Zc}.

\begin{figure}[ht]
\includegraphics[height=4.2cm]{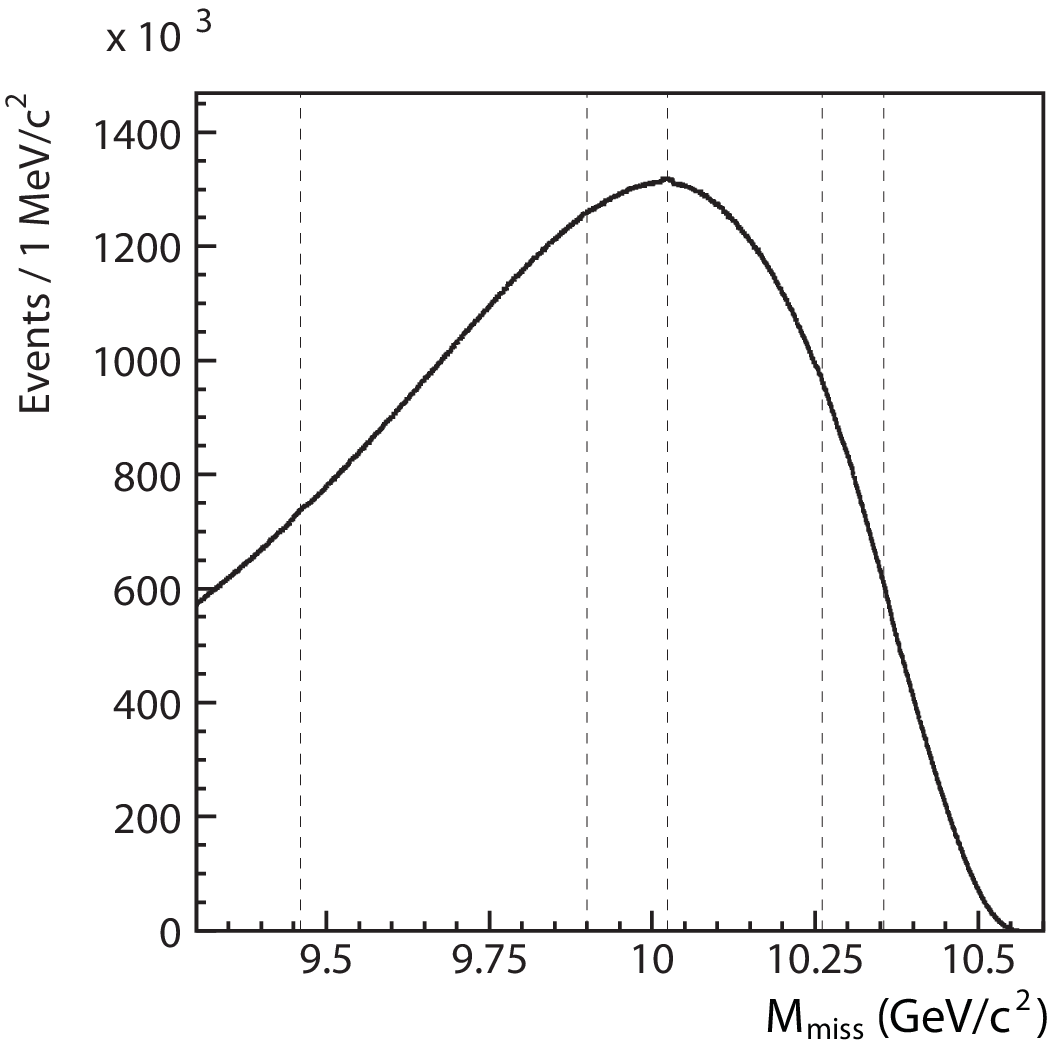}
\includegraphics[height=4.2cm]{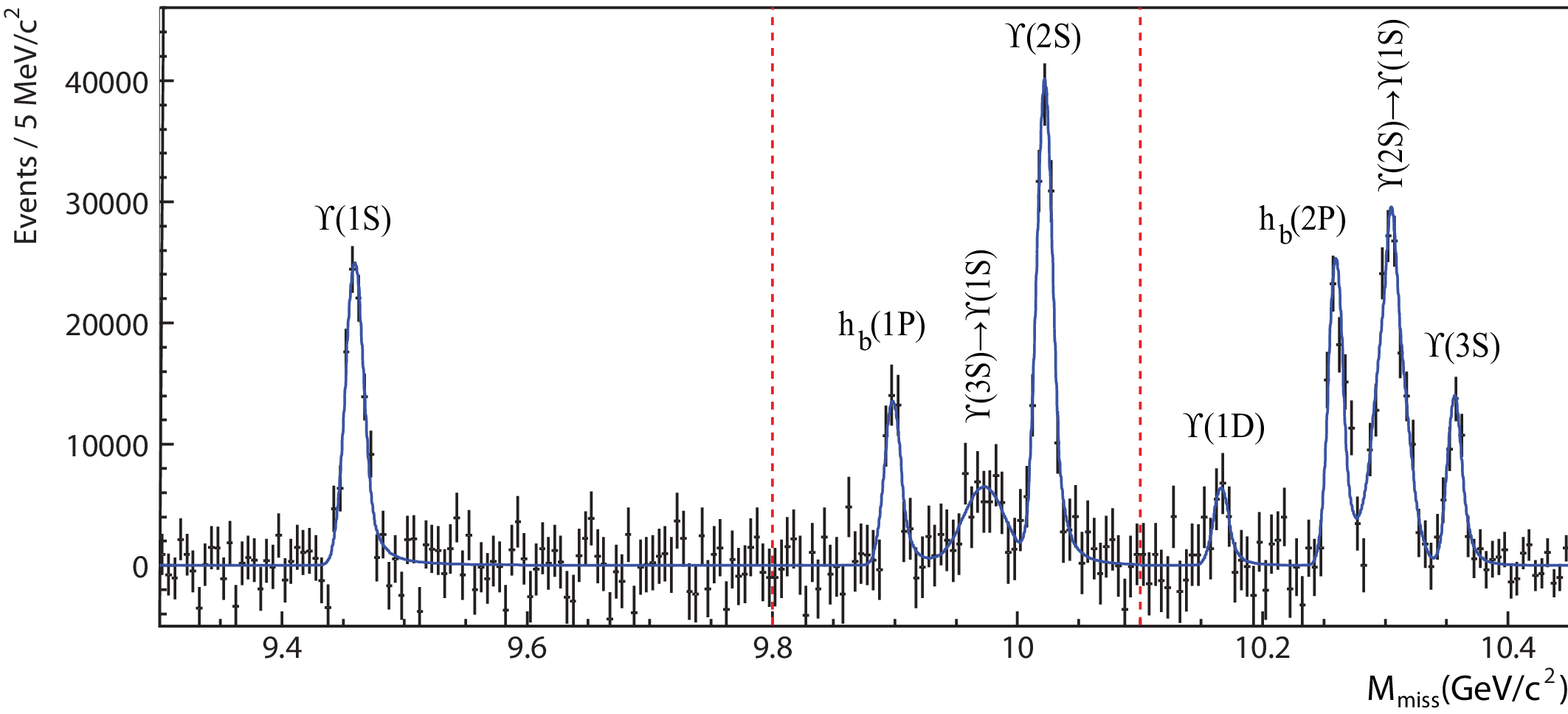}
\caption{Distribution in dipion missing mass $MM_{\pi\pi}$, (left) raw and (right) background-subtracted spectra.
\label{fig:hb}}
\end{figure}

The $h_b$ is expected to be produced through the dipion transition $\Upsilon$(5S)$\to h_b{\rm (nS)}\pi^+\pi^-$.  As there are no clear and efficient  exclusive decays expected of $h_b$, this search is conducted by taking pairs of oppositely charged pions and observing the distribution in missing mass.
The raw sample of candidates is huge but, given the high statistics, it is possible to fit to peaks with smooth polynomial background functions to identify several new states.
The raw and background-subtracted distributions are shown in Figure~\ref{fig:hb}.
The fit yields a mass of $9898.25\pm 1.06^{+1.03}_{-1.07}$MeV/$c^2$ ($10259.76\pm 0.64^{+1.43}_{-1.03}$MeV/$c^2$) for $h_b$(1P) ($h_b$(2P)), with significance  5.5$\sigma$  (11.2$\sigma$) including systematics.
This constitutes the first observation of these two states.
In the absence of hyperfine interactions, the mass of $h_b$(nP) is equal to the spin-weighted average of the corresponding $\chi_b$(nP) states, $M_0(h_b({\rm nP}))\equiv [M(\chi_{b0}({\rm nP}))+3M(\chi_{b1}({\rm nP}))+5M(\chi_{b2}({\rm nP}))]/9$,
and for heavy quarkonia the difference $\Delta M_{\rm n}=M_{meas}(h_b({\rm nP}))-M_0(h_b({\rm nP}))$ is expected to be small.
We find $\Delta M_1=1.62\pm 1.52$MeV/$c^2$ and $\Delta M_2=0.48^{+1.57}_{-1.22}$MeV/$c^2$, both consistent with zero.
The ratio of the corresponding $h_b$ and $\Upsilon$ rates, 
$R_{n}\equiv \frac{\Gamma(\Upsilon({\rm 5S})\to h_b(\rm nS)\pi^+\pi^-)}{\Gamma(\Upsilon({\rm 5S})\to \Upsilon(\rm {2S})\pi^+\pi^-)}$,
is na\"{i}vely expected to be small, as the transition to $h_b$ requires a heavy quark to flip spin, whereas that to $\Upsilon$ does not.
The observed ratios, $R_1 = 0.407\pm 0.079^{+0.048}_{-0.076}$ and
$R_2=0.78\pm 0.09^{+0.22}_{-0.10}$, possibly indicate a more complicated picture.

\begin{figure}[ht]
\includegraphics[width=4.0cm]{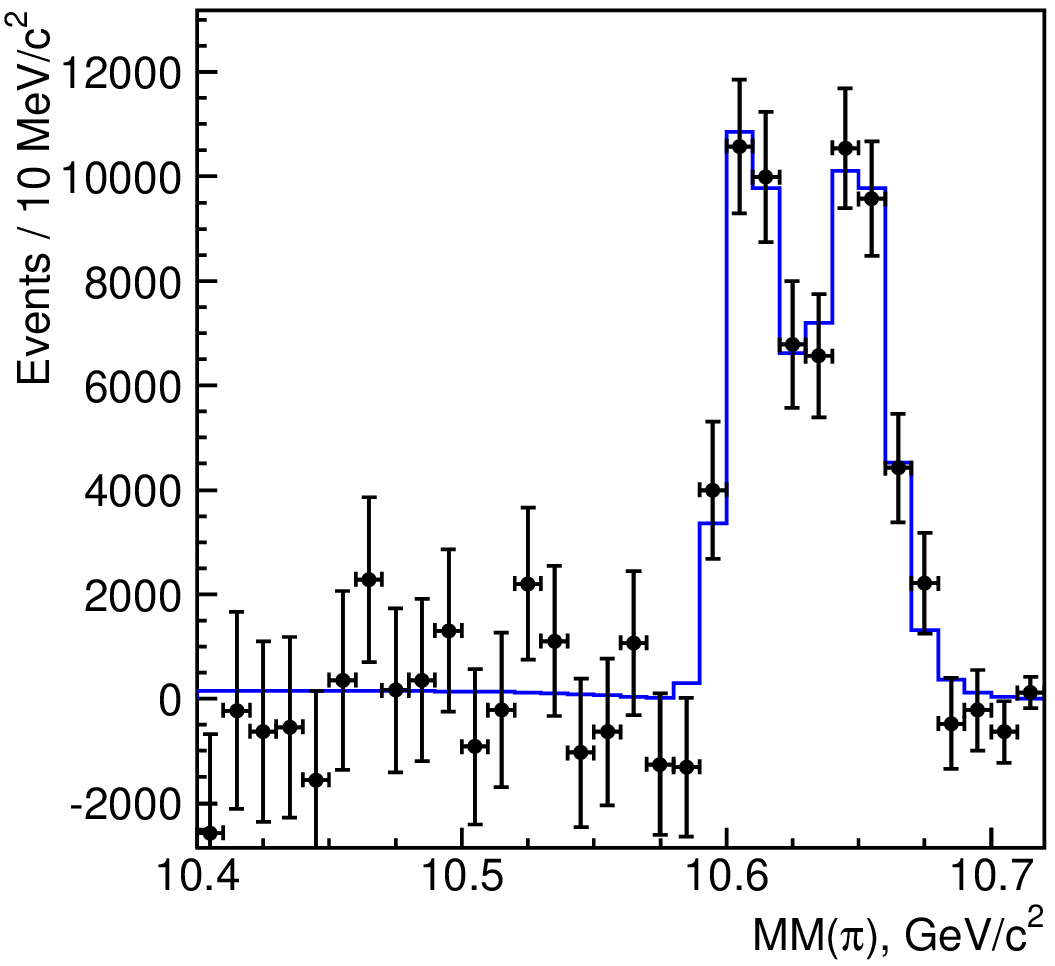}
\includegraphics[width=4.0cm]{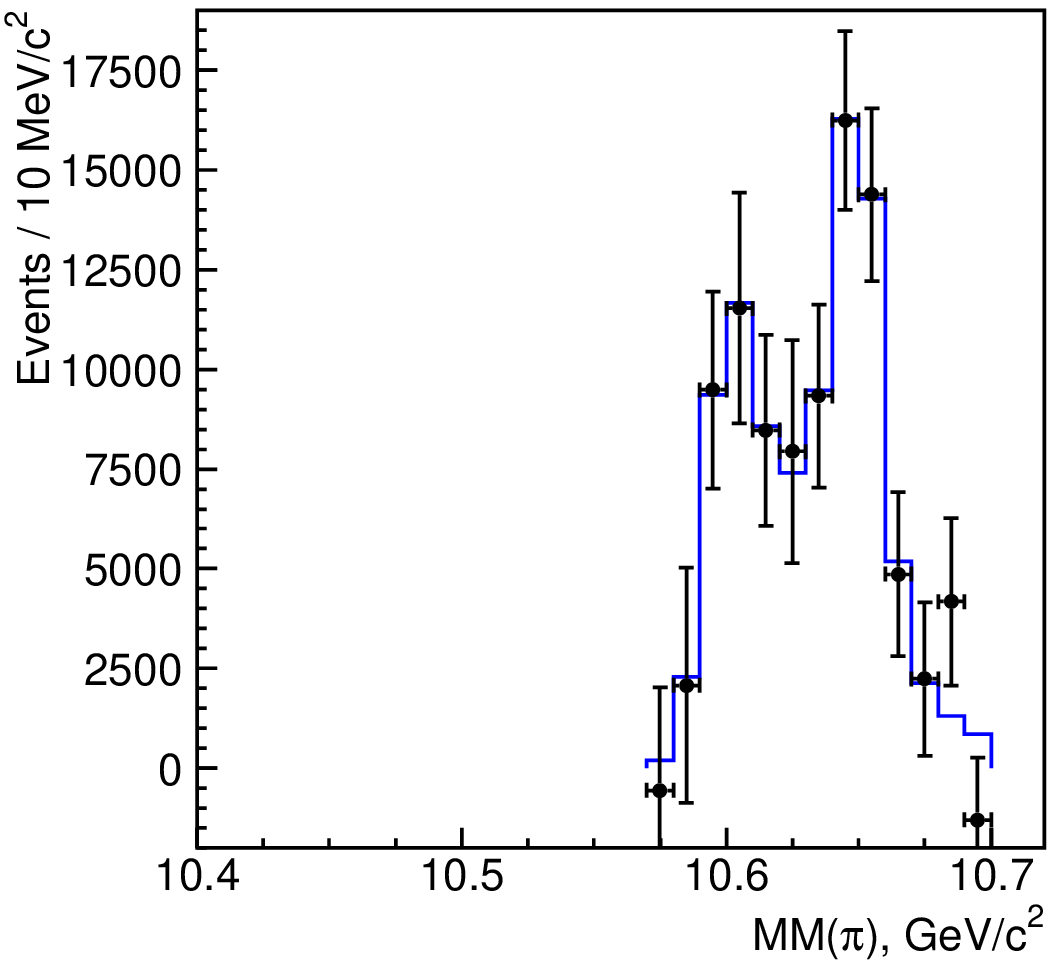}
\caption{$MM_{\pi}$ distribution of yields in (left) $h_b$(1P) and (right) $h_b$(2P).
\label{fig:mmpi}}
\end{figure}

Motivated by this and other anomalies listed above, we have studied the resonant substructure of the $\Upsilon$(5S)$\to h_b({\rm nP})\pi^+\pi^-$ and $\Upsilon$(5S)$\to \Upsilon({\rm nP})\pi^+\pi^-$ transitions.
For $\Upsilon$(5S)$\to h_b({\rm nP})\pi^+\pi^-$ we select $h_b$(nP) candidates as above and for each of the two pions construct a $h_b$(nP)$\pi$ mass via missing mass of the remaining pion, $MM(\pi)$.
The $h_b$ fit is repeated in bins of $MM(\pi)$, and the resulting yields are plotted as shown in Figure~\ref{fig:mmpi}.
Evident in both plots is a double peak structure, which is fitted to a sum of two S-wave Breit-Wigner forms plus a nonresonant background:
$|BW(s,M_1,\Gamma_1)+ae^{i\phi}BW(s,M_2,\Gamma_2)+be^{i\psi}|^2\frac{qp}{\sqrt{s}}$.
The mass values and widths for the $h_b$(1P)$\pi$ and $h_b$(2P)$\pi$ channels are in good agreement with each other (Table~\ref{tab:hbpi}). The level of non-resonant contribution is consistent with zero.

\begin{table}[h]
\begin{tabular}{c|cc| ccc| c}
\hline\hline\\
 & $h_b({\rm 1P})\pi^\pm\pi^\mp$ & $h_b({\rm 2P})\pi^\pm\pi^\mp$ & $\Upsilon({\rm 1S})\pi^\pm\pi^\mp$ &$\Upsilon({\rm 2S})\pi^\pm\pi^\mp$ &$\Upsilon({\rm 3S})\pi^\pm\pi^\mp$ & Average \\ \hline
$M_1$ (${\rm MeV}/c^2$) & 
$10605.1\pm 2.2^{+3.0}_{-1.0}$ & 
$10596\pm 7^{+5}_{-2}$ &$10609\pm 3\pm 2$&$10616\pm 2^{+3}_{-4}$&$10608\pm 2^{+5}_{-2}$&$10608\pm 2.0$ \\
$\Gamma_1$ (${\rm MeV}$) & $11.4{^{+4.5}_{-3.9}}{ ^{+2.1}_{-1.2}}$ & 
$16{^{+16}_{-10}}{ ^{+13}_{-14}}$ & $22.9\pm 7.3\pm 2$&$21.1\pm 4^{+2}_{-3}$&$12.2\pm 1.7\pm 4$&$15.6\pm 2.5$\\
$M_2$ (${\rm MeV}/c^2$) & 
$10654.5 \pm 2.5^{+1.0}_{-1.9}$ & 
$10651 \pm 4\pm 2$&$10660\pm 6\pm 2$&$10653\pm 2\pm 2$&$1-652\pm 2\pm 2$& $10653\pm 1.5$ \\
$\Gamma_2$ (${\rm MeV}$) &  
$20.9{^{+5.4}_{-1.7}}{ ^{+2.1}_{-5.7}}$ &
$12{^{+11}_{-9}}{ ^{+8}_{-2}}$ &$12\pm 10\pm 3$&$16.4\pm 3.6^{+4}_{-6}$&$10.9\pm 2.6^{+4}_{-2}$ & $14.4\pm 3.2$\\
$\phi$ ($^\circ$)& 
$188{^{+44}_{-58}}{ ^{+4}_{-9}}$ &
$255{^{+56}_{-72}}{ ^{+12}_{-183}}$&$53\pm 61^{+5}_{-50}$&$-20\pm 18^{+14}_{-9}$&$6\pm 24^{+23}_{-59}$& -- \\
\\ \hline\hline
\end{tabular}
\caption{Masses, widths, and relative phases of peaks observed in $h_b\pi$ and $\Upsilon\pi$ channels,  from fits described in text.\label{tab:hbpi}}
\end{table}

\begin{figure}[ht]
\includegraphics[width=4.0cm]{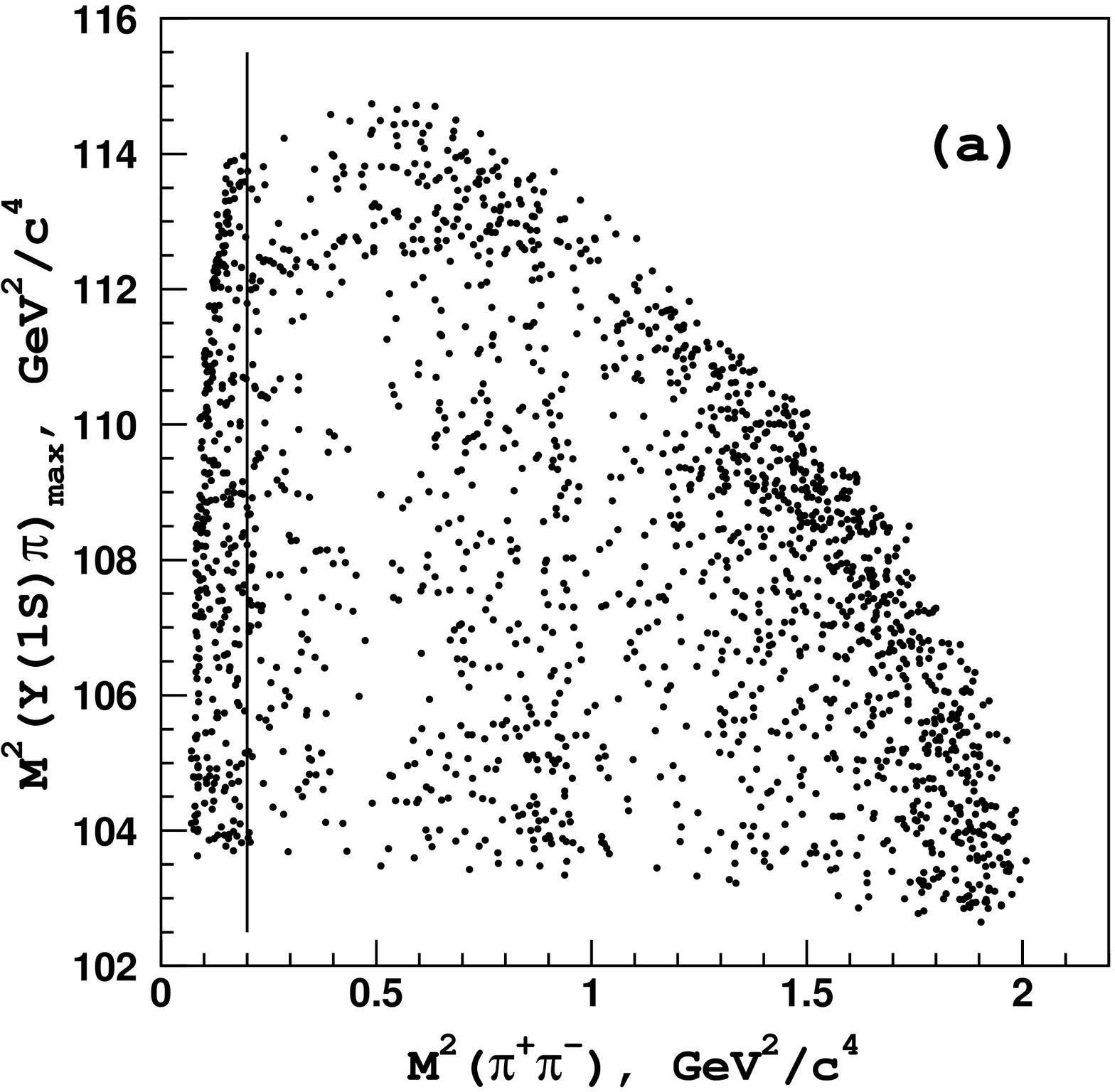}
\includegraphics[width=4.0cm]{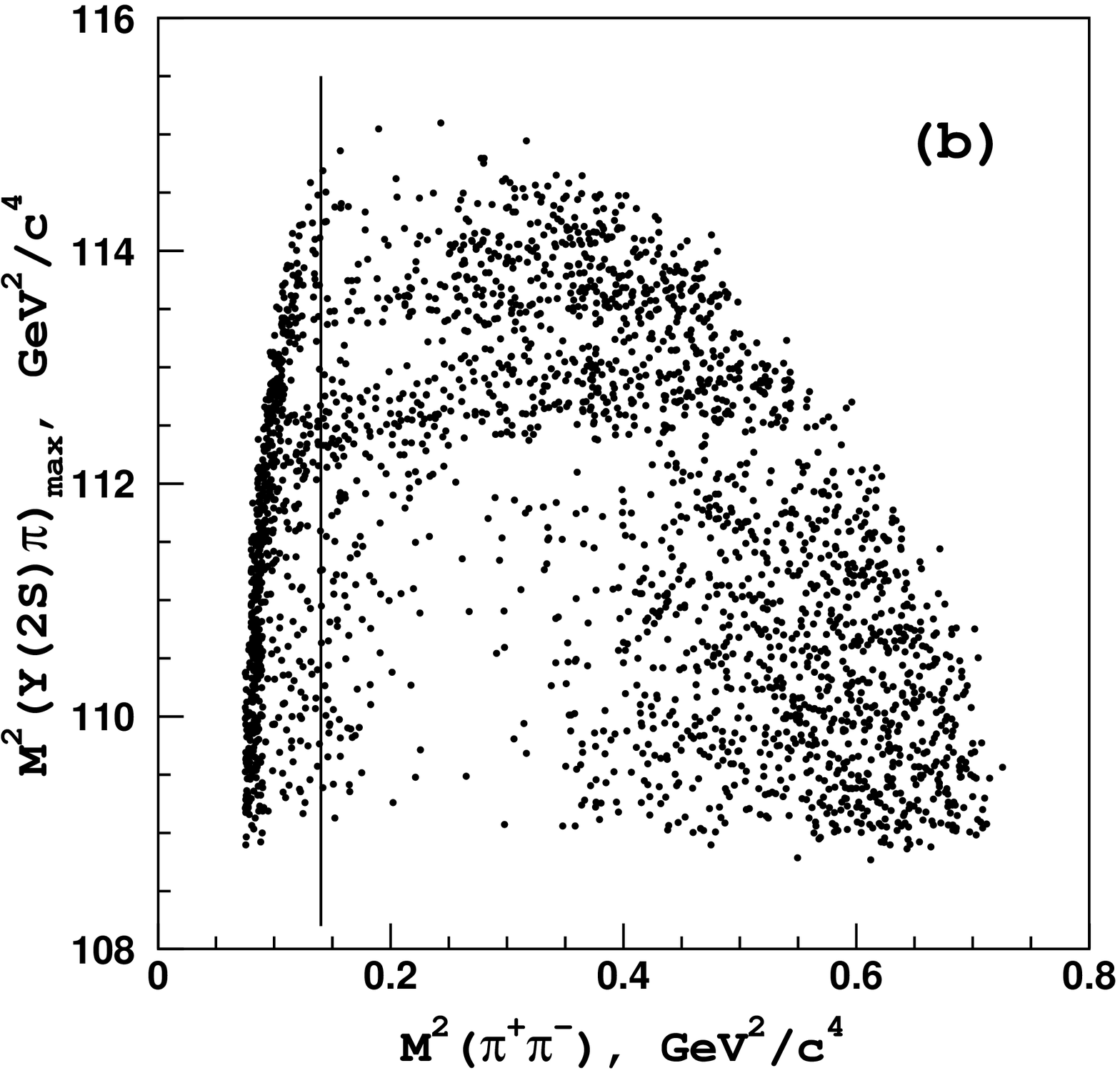}
\includegraphics[width=4.0cm]{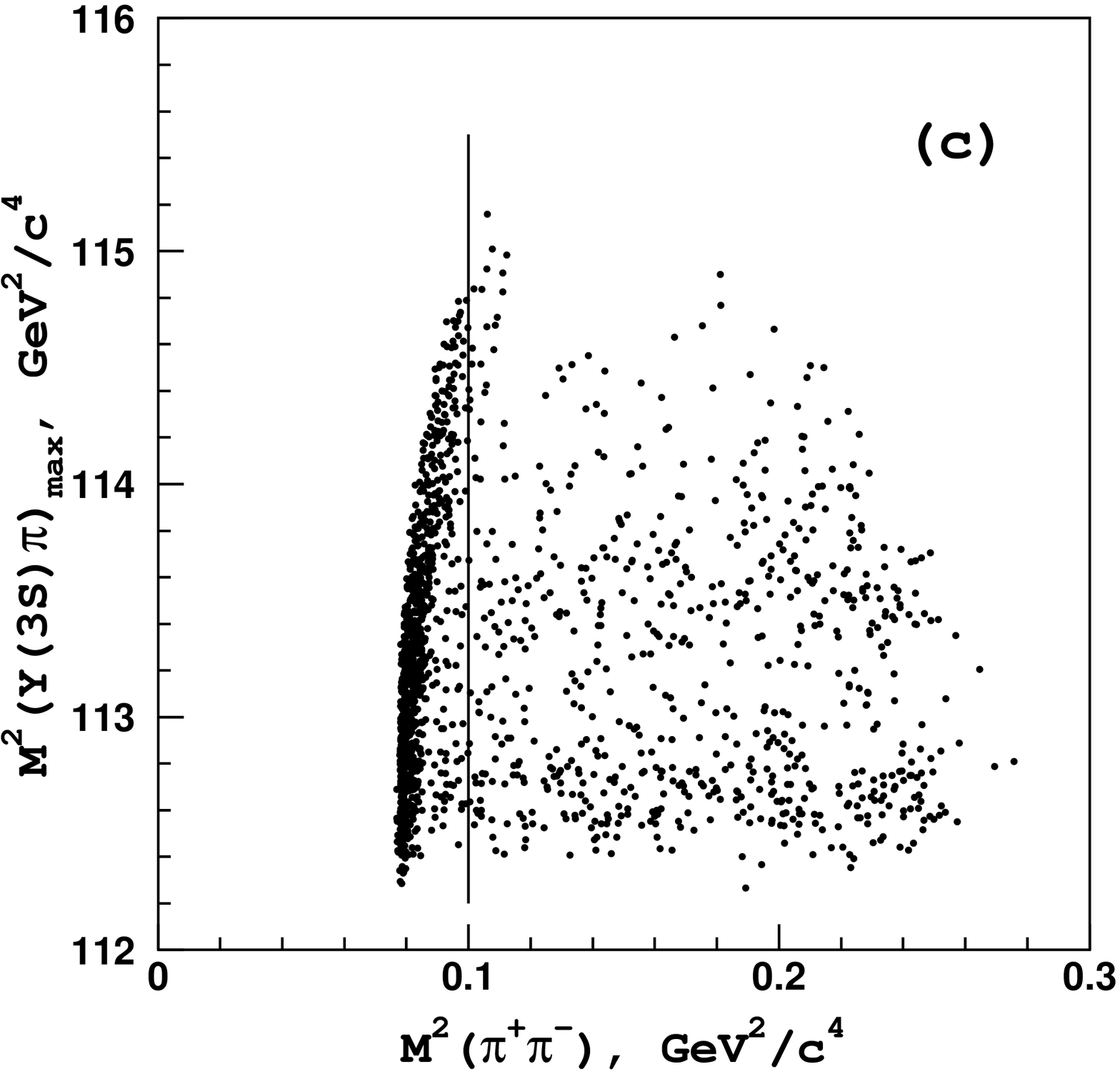}
\caption{Distribution in $M^2(\Upsilon$(nS)$\pi\ vs. M^2(\pi\pi)$ for (left) $\Upsilon$(1S)$\pi^+\pi^-$, (center) $\Upsilon$(2S)$\pi^+\pi^-$, and (right) $\Upsilon$(3S)$\pi^+\pi^-$ events.
\label{fig:YpipiDalitz}}
\end{figure}

To examine $\Upsilon$(5S)$\to \Upsilon({\rm nS})\pi^+\pi^-$ (where n=1,2,3), we collect a sample of exclusive events $\Upsilon({\rm 5S})\to\Upsilon({\rm nS})\pi^\pm\pi^\mp\{ \Upsilon({\rm nS})\to \mu^+\mu^-\}$, which is substantial and very clean, and subject it to Dalitz analysis.
The distributions of $\Upsilon({\rm nS})\pi^+\pi^-$ events in the Dalitz variables $M^2(\pi^+\pi^-)$ and $M^2(\Upsilon({\rm nS})\pi^\pm)$ are shown in Figure~\ref{fig:YpipiDalitz}.
A visual examination reveals two horizontal bands in each of the plots.
Each 2-dimensional distribution, excluding a zone of low $M^2(\pi^+\pi^-)$ (shown in the figure) to eliminate initial state radiation with photon conversion $e^+e^-\to \gamma \ell^+\ell^-\{\gamma\to e^+e^-\}$,
is fitted via unbinned maximum likelihood method to a function that includes two Breit-Wigner amplitudes $A_{Z_{b1,2}}$, a nonresonant form
$A_{\rm NR} = c_1+c_2M^2_{\pi\pi}$\cite{voloshin}, and the $\pi\pi$ resonances $f_0$(980) and $f_2$(1275):
$S=|A_{Z_{b1}}+A_{Z_{b2}}+A_{\rm NR}+A_{f_0(980)}+A_{f_2(1275)}|^2$.
The fitted masses, widths, and relative phases are shown in Table~\ref{tab:hbpi}.
The measured masses and widths for all five channels are in good agreement, and the overall averages are included in the table.  
It is noted that the relative phases of the $h_b$ ($\Upsilon$) channels are consistent with $180^\circ$ ($0^\circ$) and that the two masses are just above threshold for $B_s^*\bar B_s$ and $B_s^*\bar B_s^*$, respectively, characteristics predicted for "meson molecule" models\cite{molecule}.
This constitutes the first observation of two charged bottomonium-like states, which we shall call $Z_b(10610)$ and $Z_b(10650)$.

In summary the Belle experiment, which was designed to  measure $CP$-asymmetry in $B$ decay, is exploring the $\Upsilon$(5S) region and reports first results based on 121.4~fb$^{-1}$ of data collected on the $\Upsilon$(5S) resonance.
We have made the first observation of a baryonic $B_s$ decay, in the mode $B_s\to \Lambda_c^+\pi^-\bar\Lambda$.
Using full reconstruction, we identify $\Upsilon$(5S)$\to B^{(*)0}B^{(*)-}\pi^+$ events to tag $B^0$ flavor and make a time-independent measurement of $\sin 2\phi_1$.
We have also made the first observation of the $h_b$(1P) and $h_b$(2P) as well as two new charged bottomonium-like states, $Z_b^+(10610)$ and $Z_b^+(10650)$, which are each observed in five different channels.

\section*{Acknowledgments}
The author wishes to thank the organizers and staff of  PANIC.
This work is supported by Department of Energy grant \# DE-FG02-84ER40153.

\bibliographystyle{aipproc} 

\end{document}